\documentclass[aps,prb,showpacs,twocolumn]{revtex4}
\usepackage{bm}
\usepackage[dvips]{graphicx,color}

\begin{document}

\title{Phonons in a nanoparticle mechanically coupled to a substrate}

\author{Kelly R. Patton and Michael R. Geller}

\affiliation{Department of Physics and Astronomy, University of Georgia,
 Athens, Georgia 30602-2451}

\date{\today}

\begin{abstract}
The discrete nature of the vibrational modes of an isolated nanometer-scale
solid dramatically modifies its low-energy electron and phonon dynamics
from that of a bulk crystal. However, nanocrystals are usually
coupled---even if only
weakly---to an environment consisting of other nanocrystals, a support
matrix, or a solid substrate, and this environmental interaction will modify
the vibrational properties at low frequencies. In this paper we investigate
the modification of the vibrational modes of an insulating spherical
nanoparticle caused by a weak {\it mechanical} coupling to a semi-infinite
substrate. The phonons of the bulk substrate act as a bath of harmonic 
oscillators, and the coupling to this reservoir shifts and broadens
the nanoparticle's modes. The vibrational density of states in the nanoparticle
is obtained by solving the Dyson equation for the phonon propagator, and we 
show that  environmental interaction is especially important at low
 frequencies. As a probe of the modified phonon
spectrum, we consider nonradiative energy relaxation of a localized
 electronic impurity state in the nanoparticle, for which good agreement with
experiment is found.
\end{abstract}


\maketitle

\section{Introduction}
\label{Introduction}

There is currently great interest in properties of nanometer-scale mechanical
systems, such as cantilevers, nanoparticles, and resonators.\cite{roukes}
Because of the extremely small size and volume-to-surface ratio of these
 systems, 
interactions with their surroundings can dramatically
alter their properties.  In particular, it is well known that the vibrational
spectrum of 
an isolated nanometer-scale crystal, being discrete, is qualitatively 
different than 
that of the same bulk material, leading to important changes in any property
dependent on the phonon density of states (DOS).  The differences 
between the vibrational DOS in a nanoparticle and a bulk solid
are most evident at low frequencies: A spherical nanoparticle 
with diameter $d$ and characteristic bulk sound velocity $v$ can not support
 a mode
with frequency less than about $2\pi v/d$. Thus, an acoustic ``gap'' in the
low-energy  phonon spectrum is present in contrast with  that of the bulk, 
which has 
a continuous spectrum down to zero energy.  However, mechanical interaction
with the environment will modify the discrete nature of the modes.

In an interesting experiment by Yang {\it et al.},\cite{yang} the phonon
DOS deep inside this gap was measured in insulating $\rm Y_{2}O_{3}$
nanoparticles. The experiment consisted of nanoparticles whose sizes
ranged from 7 to 23 nm in diameter and was performed by measuring
the nonradiative lifetimes of an excited electronic state of a 
$\rm Eu^{3+}$ dopant. The lowest supported mode or Lamb mode for 
nanoparticles of these sizes is about $10\,{\rm cm}^{-1}$.  At 
$3\,{\rm cm}^{-1}$ the DOS measured was more than 100 times smaller than 
that of bulk $\rm Y_{2}O_{3}$ (at 3 cm$^{-1}$).

In this paper we propose and investigate a mechanism that could
be responsible for 
the observed broadening of the nanoparticles' phonon modes. Several possible
broadening mechanisms could be 
responsible for the observed effect. For example, anharmonicity leads to
broadening and, therefore, to a low-energy DOS, but anharmonicity
 is ineffective
at low energy and was found to be too small to account for the 
experiment.\cite{vadim}  Another possibility could be adsorbed ``dirt''
on the outside of
the nanoparticle.  This might lower the Q factor of the nanoparticle,
regarding it as a
resonator, reflecting a broadening of the vibrational modes. A third, and in
our opinion more likely mechanism, follows from the 
realization that these nanoparticles are not isolated, but instead were 
prepared
in a powered form. Thus, each nanoparticle is in weak contact with a 
cluster of other nanoparticles. Because the cluster is relatively large, 
including several hundred nanoparticles, each nanoparticle  is 
mechanically 
coupled to a reservoir that has a continuous vibrational DOS at low energy.  
This 
interaction broadens the modes and allows  phonons in the nanoparticle
to escape and be absorbed into the cluster. We investigate the affect
this mechanical environmental interaction has on the nanoparticle's phonon DOS.

Because we are only interested in determining the correct origin 
of broadening,
and do not hope to be able to exactly reproduce the experimental results
of Ref.~\onlinecite{yang}, we propose the
following simplified model:  The cluster of nanoparticles is replaced by 
a semi-infinite elastic substrate, and one nanoparticle is placed in weak 
mechanical contact with it.
The weak contact is imagined to be a few atomic bonds or small
neck of material, which we model by a harmonic spring.  For 
simplicity, we take the substrate and the nanoparticle to be made out of
the same isotropic elastic material.  Because we are interested in the 
low energy regime,
continuum elasticity theory will be used to describe the nanoparticle and the 
substrate. After defining and analyzing our simplified model, in Sections 
\ref{golden rule lifetimes} through  \ref{results}, we explain in Section 
\ref{comparison with experiment} how the model can be adapted to address the 
experiment of Ref.~\onlinecite{yang}, and good agreement is obtained.

The simple model we study is related to, but different than, models used
to study energy relaxation by molecules adsorbed on surfaces.\cite{zangwill}
However, in surface science the interest is usually in the relaxation of rigid
translational motion, rotational motion, or simple internal vibrations of 
adsorbates. In contrast, we investigate the broadening of complex internal 
vibrational modes of much larger objects (which are crystalline). Also, our
work has much in common with that of Gurevich and Schober,\cite{g and s} where
many of the same considerations and modeling were used to study the Lamb-mode
 decay rate, of nanoparticles caused by both anharmonicity and coupling to an 
enviroment of other nanoparticles.

\section{NANOPARTICLE AND SUBSTRATE MODEL}
\label{naonoparticle and substrate section}
\begin{figure}
\includegraphics[width=8.5cm]{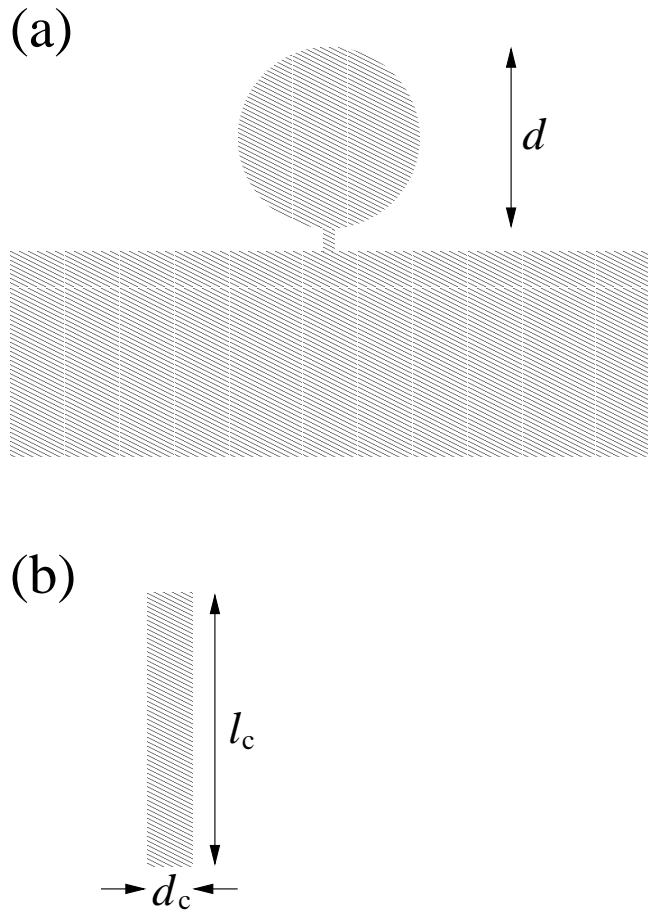}
\caption{\label{model}(a) Model of nanoparticle, substrate, and connecting region. (b) Expanded view of connecting region with dimensions $d_{\rm c}$ and $l_{\rm c}$}
\end{figure}

As mentioned in the introduction, the model we  study is that of a 
single nanoparticle in weak mechanical contact with a semi-infinite substrate.
Linear elasticity theory will be used to describe the phonons of this system.
We assume the nanoparticle and substrate to be made of an isotropic non-polar
material.
Because we take the nanoparticle and substrate to be made of the same
material, we will use the same density $\rho$ and Lam\'e coefficients
$\lambda$ and $\mu$ for both.  The Lagrangian for the entire system is given
by 
\begin{equation}
L=\int_{\sf V} d^3r\left[{\textstyle{1 \over 2}} \rho \, (\partial_t {\bf u})^2 - {\textstyle{1 \over 2}}
\lambda \, u_{ii}^2 - \mu \, u_{ij}^2\right],
\label{original lagrangian} 
\end{equation}
where ${\bf u}({\bf r},t)$ is the displacement field, and
\begin{equation}
u_{ij} \equiv (\partial_i u_j + \partial_j u_i)/2
\end{equation}
is the strain tensor. ${\sf V}$ is the volume of the nanoparticle, substrate,
and connecting material, as shown in Fig.~\ref{model}.
Because the Lagrangian density
is local, the integration volume in (\ref{original lagrangian}) can be split 
into three independent parts:
 the nanoparticle, the substrate, and the connecting region. In the limit of
weak coupling (diameter $d_{\rm c}$ of connecting region much smaller than 
$d$), the surface area on the nanoparticle and substrate over which the actual
boundary conditions differs from stress-free conditions are negligible, and the Hamiltonian can be written as
\begin{equation}
H=H_{\rm nano}+H_{\rm sub}+\delta H,
\end{equation}
where $H_{\rm nano}$ is the Hamiltonian for an isolated nanoparticle (with
stress-free boundaries),
$H_{\rm sub}$ is that for an isolated substrate, and $\delta H $ is
the interaction between the two.  The connecting region is taken to be a 
few atomic
bonds or small neck of material, as shown in Fig.~\ref{model}. 
We further approximate this mechanical coupling to be a harmonic spring
potential
\begin{equation}
\delta H =\frac{1}{2}K:\left[u^z_{\rm nano}({\bf r}_{0})-u^z_{\rm sub}({\bf r}_{0})\right]^2:,
\label{perturbation}
\end{equation}
where $K$ is an effective spring constant, and $u^z_{\rm nano}$ 
and $u^z_{\rm sub}$ are the $z$ components of the displacement
field of the nanoparticle and substrate at the point of contact, 
${\bf r}_{0}$. We take the $z$ direction to be along the upward pointing 
normal to the substrate surface.  The Hamiltonians we will introduce below 
for $H_{\rm nano}$
and $H_{\rm sub}$ are normal-ordered; therefore, it is necessary to normal 
order
$\delta H$ as well.  This operation is denoted by the colons in 
Eq.~(\ref{perturbation}).

Our analysis will require the vibrational normal modes and spectra of
the isolated nanoparticle and semi-infinite substrate, calculated with
stress-free boundary conditions.  The long-wavelength modes of interest
here may be obtained from elasticity theory, to which we now turn.

\subsection{Isolated Nanoparticle}
Here we derive the normal modes of an isolated  elastic sphere. 
The method we shall use is different than 
(but equivalent to) that used in the classic paper by Lamb,\cite{lamb} but
is better suited for our purposes.
 The equation of motion of (\ref{original lagrangian}) is 
\begin{equation}
\partial_t^2 {\bf u} - v_{\rm l}^2 \mbox{\boldmath $\nabla$} (\mbox{\boldmath $\nabla$}
\cdot {\bf u}) +  v_{\rm t}^2 \mbox{\boldmath $\nabla$} \times \mbox{\boldmath $\nabla$}
\times {\bf u} = 0,
\label{equation of motion}
\end{equation}
where $v_{\rm l} \equiv \sqrt{ (\lambda + 2 \mu) / \rho}$ is the bulk
longitudinal
sound velocity and $v_{\rm t} \equiv \sqrt{\mu / \rho}$ is the transverse 
velocity.
To solve Eq.~({\ref{equation of motion}})
the displacement field can be decomposed into longitudinal and 
transverse parts,
\begin{equation}
{\bf u} = {\bf u}_{\rm l} + {\bf u}_{\rm t},
\label{decomposition}
\end{equation}
where
\begin{equation}
 \mbox{\boldmath $\nabla$} \times{\bf u}_{\rm l} =0
\end{equation}
and
\begin{equation}
\mbox{\boldmath $\nabla$} \cdot {\bf u}_{\rm t} = 0.
\end{equation}
With harmonic time dependence, the equation of motion Eq.~({\ref{equation of motion}}) then separates into two
vector Helmholtz equations for the longitudinal and transverse parts,
\begin{equation}
\big(\nabla^2 + p^2 \big) {\bf u}_{\rm l} = 0, \ \ \ \ \ \ \ \ \ p \equiv \omega/v_{\rm l}
\label{longitudinal wave equation}
\end{equation}
and
\begin{equation}
\big(\nabla^2 + q^2 \big) {\bf u}_{\rm t} = 0, \ \ \ \ \ \ \ \ \ q \equiv \omega/v_{\rm t}.
\label{transverse wave equation}
\end{equation}
The longitudinal equation (\ref{longitudinal wave equation}) can be solved by 
introducing a scalar potential
\begin{equation}
{\bf u}_{\rm l} =  \mbox{\boldmath$\nabla$} \phi^{(p)},
\label{scalar potential}
\end{equation}
where $\phi^{(p)}$ is a solution of the scalar Helmholtz equation $\big( \nabla^2 + p^2 \big) \phi^{(p)} =0$.
The transverse equation (\ref{transverse wave equation}) has two linearly
independent 
solutions, ${\bf u}_{\rm t} = {\bf M} \ {\rm and} \ {\bf N},$ where
\begin{equation}
{\bf M} = \mbox{\boldmath$\nabla$} \phi^{(q)} \times {\bf r}
\label{M definition}
\end{equation}
and
\begin{equation}
{\bf N} = {\textstyle{1 \over q}} \mbox{\boldmath$\nabla$} \times {\bf M}.
\label{N definition}
\end{equation}
Here $\phi^{(q)}$ is a solution of $\big( \nabla^2 + q^2 \big) \phi^{(q)} =0$.
The prefactor $1/q$ is included for dimensional
 convenience.
The scalar Helmholtz equations are separable in spherical coordinates
 and the solutions can 
be written as
\begin{equation}
\phi_{lm}({\bf r}) \equiv j_l(kr) \, Y_{lm}(\theta,\varphi),\ \ \ k=p,q
\label{scalar Helmholtz solutions}
\end{equation}
where
\begin{equation}
j_l(x) \equiv \sqrt{\textstyle{\pi \over 2 x}} \ J_{l + {1 \over 2}}(x)
\label{spherical Bessel function definition}
\end{equation}
is a spherical Bessel function of the first kind (regular at origin) and
\begin{equation}
Y_{lm}(\theta,\varphi) \equiv (-1)^m \sqrt{ {2l + 1 \over 4 \pi} \, {(l-m)! \over (l+m)!}}
\ P_{lm}\big(\! \cos \theta \big) \, e^{i m \varphi}.
\label{spherical harmonic definition}
\end{equation}
Here 
\begin{equation}
P_{lm}(x) \equiv (1-x^2)^{m \over 2} \, {\partial^m \over \partial x^m} \, P_l(x),
\end{equation}
where $P_{l}(x)$ are Legendre polynomials.

Now we use the $\phi_{lm}$ to construct three linearly independent 
solutions of 
(\ref{equation of motion}),
\begin{eqnarray}
{\bf L}_{lm} &\equiv& {\textstyle{1 \over p}} \mbox{\boldmath$\nabla$} 
\phi_{lm}(pr), \\
{\bf M}_{lm} &\equiv& \mbox{\boldmath$\nabla$} \phi_{lm}(qr) \times {\bf r}, \\
{\bf N}_{lm} &\equiv& {\textstyle{1 \over q}}  \mbox{\boldmath$\nabla$} \times 
{\bf M}_{lm}.
\end{eqnarray}
The general solution is a linear combination of ${\bf L}_{lm}$,
 ${\bf M}_{lm}$, and
${\bf N}_{lm}$,
\begin{equation}
{\bf u}({\bf r}) = \sum_{lm} \big[ a_{lm}{\bf L}_{lm} +b_{lm}{\bf M}_{lm}+c_{lm}{\bf N}_{lm} \big].
\label{general solution expansion}
\end{equation} Although they are linearly independent, the vector 
fields ${\bf L}_{lm}$, 
${\bf M}_{lm}$, and ${\bf N}_{lm}$ are not orthogonal in space. However, 
they  can be written in terms of orthogonal vector spherical harmonics 
${\bf P}_{lm},\ {\bf B}_{lm}$, and ${\bf C}_{lm}$, defined as
\begin{equation}
{\bf P}_{lm}(\Omega)\equiv Y_{lm}(\Omega){\bf e}_{r}
\end{equation}
\begin{equation}
{\bf B}_{lm}(\Omega)\equiv \frac{1}{\sqrt{l(l+1)}}\left(\partial_{\theta}Y_{lm}(\Omega){\bf e}_{\theta}+\frac{imY_{lm}(\Omega)}{\sin \theta}{\bf e}_{\varphi}
\right)
\end{equation}

\begin{equation}
{\bf C}_{lm}(\Omega)\equiv \frac{1}{\sqrt{l(l+1)}}\left(\frac{imY_{lm}(\Omega)}{\sin \theta}{\bf e}_{\theta}-\partial_{\theta}Y_{lm}(\Omega){\bf e}_{\varphi}\right),
\end{equation}
with the following properties
\begin{equation}
\int d\Omega\ {\bf X}^{*}_{lm}\cdot{\bf X}_{l'm'}=\delta_{ll'}\delta_{mm'}
\end{equation}
for ${\bf X}\ \in\ {\bf B},{\bf C},{\bf P}$ and
\begin{equation}
\int d\Omega\ {\bf X}^{*}_{lm}\cdot{\bf X}'_{l'm'}=0
\end{equation}
for ${\bf X}\neq {\bf X}'$.
Expressed in terms of vector spherical harmonics, ${\bf L}_{lm},\ {\bf M}_{lm}$ and ${\bf N}_{lm}$ are given by
\begin{equation}
{\bf L}_{lm}=j'_{l}(pr){\bf P}_{lm}(\Omega)+\frac{\sqrt{l(l+1)}}{pr}j_{l}(pr){\bf B}_{lm}(\Omega),
\end{equation}
\begin{eqnarray}
{\bf M}_{lm}=\sqrt{l(l+1)}j_{l}(qr){\bf C}_{lm}(\Omega),
\end{eqnarray}
and
\begin{eqnarray}
{\bf N}_{lm}&=&\frac{l(l+1)}{qr}j_{l}(qr){\bf P}_{lm}(\Omega)+\frac{\sqrt{l(l+1)}}{qr}\nonumber\\&\times&\Big[j_{l}(qr)+qrj'_{l}(qr)\Big]{\bf B}_{lm}(\Omega),
\end{eqnarray}
where prime denotes differentiation with respect to the argument.

Next we impose stress-free
boundary conditions
\begin{equation}
\sigma_{ij} n_j = 0
\label{stress-free boundary conditions}
\end{equation}
at the surface $r = R$ of the nanoparticle. Here ${\bf n}$ is an outward pointing normal
vector and $\sigma_{ij}$ is the strain tensor. In an isotropic elastic continuum,
\begin{equation}
\sigma_{ij} = \lambda \, (\mbox{\boldmath$\nabla$} \cdot {\bf u}) \, \delta_{ij} + 2 \mu \, 
u_{ij}.
\label{Hookes law}
\end{equation}
In spherical coordinates (\ref{stress-free boundary conditions}) implies
\begin{equation}
\sigma_{rr} = \sigma_{\theta r} = \sigma_{\varphi r} = 0.
\label{stress tensor conditions}
\end{equation}
The three conditions (\ref{stress tensor conditions}) require that
\begin{equation}
\lambda \, (\mbox{\boldmath$\nabla$} \cdot {\bf u}) + 2 \mu \, u_{rr} = 0,
\label{r condition}
\end{equation}
\begin{equation}
u_{\theta r} = 0,
\label{theta condition}
\end{equation}
and
\begin{equation}
u_{\varphi r} = 0.
\label{phi condition}
\end{equation}

In terms of the displacement field,
\begin{eqnarray}
u_{rr} &=& \partial_r u_r, \\
u_{\theta r} &=& {\textstyle{1 \over 2}}(\partial_r u_\theta - {\textstyle{1 \over r}} \,
u_\theta + {\textstyle{1 \over r}} \, \partial_\theta u_r), \\
u_{\varphi r} &=& {\textstyle{1 \over 2}}( {\textstyle{1 \over r \sin\theta }} \,
\partial_\varphi u_r + \partial_r u_\varphi - {\textstyle{1 \over r}} \, u_\varphi).
\end{eqnarray}
The boundary condition equations (\ref{stress tensor conditions}) then become
\begin{eqnarray}
& &a_{lm}\big[-\lambda pj_{l}(pR)\,Y_{lm}+2\mu pj_{l}''(pR)\,Y_{lm}\big]\nonumber\\
& &+\:c_{lm}2\mu l(l+1)\,{\sf E}\,Y_{lm}=0, 
\label{condition 1}
\end{eqnarray}
\begin{eqnarray}
& &a_{lm}2\,{\sf D}\partial_{\theta}\,Y_{lm}+b_{lm}im\,{\sf E}\,Y_{lm}\,{\csc\theta}\nonumber\\
& &+\:c_{lm}\,{\sf F}\partial_{\theta}Y_{lm}=0, 
\label{condition 2}
\end{eqnarray}
and
\begin{eqnarray}
& &a_{lm}2im\,{\sf D}\,Y_{lm}\,{\csc\theta}+b_{lm}\,{\sf E}\partial_{\theta}Y_{lm}\nonumber\\
& &+\:c_{lm}im\,{\sf F}\,Y_{lm}{\csc\theta}=0,
\label{condition 3}
\end{eqnarray}
where
\begin{eqnarray}
&{\sf D}&\equiv \frac{j'_{l}(pR)}{R}-\frac{j_{l}(pR)}{pR^2},\\
&{\sf E}&\equiv \frac{j'_{l}(qR)}{R}-\frac{j_{l}(qR)}{qR^2},\\
&{\sf F}&\equiv qj''_{l}(qR)+\frac{l(l+1)}{qR^2}j_{l}(qR)-\frac{2j_{l}(qR)}{qR^2}.
\end{eqnarray}
Finally, we rewrite (\ref{condition 1}) though (\ref{condition 3}) in matrix
form as
\begin{widetext}
\begin{displaymath}
\left( \begin{array}{ccc}
-\lambda pj_{l}(pR)Y_{lm}+2\mu pj_{l}''(pR)Y_{lm}\vspace{.0cm} & 0\vspace{.0cm} & 2\mu l(l+1){\sf E}Y_{lm}\vspace{.0cm} \\2{\sf D}\partial_{\theta}Y_{lm}\vspace{.0cm} & im{\sf E}Y_{lm}{\csc\theta}\vspace{.0cm} &{\sf F}\partial_{\theta}Y_{lm}\vspace{.0cm} \\ 2im{\sf D}Y_{lm}{\csc\theta} & {\sf E}\partial_{\theta}Y_{lm}& im{\sf F}Y_{lm}{\csc\theta} 
\end{array} \right)\left( \begin{array}{c} a_{lm}\vspace{.0cm}\\ b_{lm}\vspace{.0cm}\\  c_{lm}\end{array}\right)=0.
\end{displaymath}
\end{widetext}
For a nontrivial solution of (\ref{condition 1}) through (\ref{condition 3}) 
to exist, the determinant of the above matrix must vanish.
Taking the determinant and simplifying we find
\begin{equation}
\big[-\lambda pj_{l}(pR)+2\mu pj_{l}''(pR)\big]{\sf E}\,{\sf F}-4\mu l(l+1){\sf E}^2\,{\sf D}=0.
\end{equation}
This implies that either
\begin{equation}
{\sf E}=\frac{j'_{l}(qR)}{R}-\frac{j_{l}(qR)}{qR^2}=0
\label{torsional condition}
\end{equation}
or
\begin{equation}
\big[-\lambda pj_{l}(pR)+2\mu pj_{l}''(pR)\big]{\sf F}-4\mu l(l+1){\sf E}\,{\sf D}=0
\label{spheroidal condition}.
\end{equation}

If ({\ref{torsional condition}}) is met, then this imposes certain constraints on $a_{lm},b_{lm}$ and $c_{lm}$, 
which require $a_{lm}=c_{lm}=0$.  This can easily be  seen in the above matrix by setting ${\sf E}=0$. 
If ({\ref{spheroidal condition}}) is met, $b_{lm}$ has to be zero.
In conclusion, we have two branches of vibrational modes. The branch in which
({\ref{torsional condition}}) is satisfied,
\begin{equation}
{\bf u}({\bf r})=b_{lmn}{\bf M}_{lmn}({\bf r}),
\end{equation}
are referred to as the {\it torsional} modes, where $n$ specifies the radial quantum
number.
The other branch is found when ({\ref{spheroidal condition}}) is satisfied,
\begin{equation}
{\bf u}({\bf r})=a_{lmn}{\bf L}_{lmn}({\bf r})+c_{lmn}{\bf N}_{lmn}({\bf r}),
\end{equation}
which are the {\it spheroidal} modes.

To quantize the vibrational modes we write the displacement field as\cite{footnote1}
\begin{equation}
{\bf u}_{\rm nano}({\bf r})=\sum _{J}\sqrt{\frac{\hbar}{2\rho \omega_{J}}}\left[a_{J}{\bf \Psi}_{J}({\bf r})+a_{J}^{\dag}{\bf \Psi}_{J}^{*}({\bf r})\right],
\label{nano phonon}
\end{equation}
where 
\begin{equation}
J=[{\rm S}\,{\rm or}\,{\rm T},n,l,m]
\end{equation}
is a label uniquely specifying a nanoparticle eigenmode.  The first entry S or
T specifies whether the mode is in the spheroidal or torsional branch,
respectively. $n$ is the radial quantum number and $l$ and $m$ are the
usual angular momentum quantum numbers.
$a$ and $a^{\dag}$ are phonon annihilation and creation operators which satisfy the Bose commutation relation
\begin{equation}
[a_{J},a_{J'}^{\dag}]=\delta_{JJ'}.
\end{equation}
The ${\bf \Psi}_{J}$ are vibrational eigenvectors normalized such that
\begin{equation}
\int_{V}|{\bf \Psi}_{J}({\bf r})|^2d^3r=1,
\end{equation}
where $V$ is the volume of the nanoparticle.
Assuming (without proof) that the modes ${\bf \Psi}_{J}$ form a complete
set, 
\begin{equation}
\sum_{J}\Psi^{i*}_{J}({\bf r})\Psi^{j}_{J}({\bf r}')=\delta^{ij}\delta({\bf r}-{\bf r}'),
\end{equation}
it can easily be shown that ${\bf u}$ satisfies the correct equal-time
canonical commutation relation with ${\bm \pi}\equiv\rho\partial_{t}{\bf u}$, namely

\begin{equation}\left[u^{i}({\bf r}), \pi^{j}({\bf r}')\right]=i\hbar\delta^{ij}\delta({\bf r}-{\bf r}').
\end{equation}
\subsection{Isolated Substrate}
The vibrational modes for a semi-infinite isotropic elastic substrate,
with a free surface
at the $xy$ plane and extending to infinity in the negative $z$ direction,
were quantized  previously by Ezawa;{\cite{ezawa}} therefore, the details 
will be left out here. The displacement field can be written as\cite{footnote1}
\begin{equation}
{\bf u}_{\rm sub}({\bf r})=\sum _{I}\sqrt{\frac{\hbar}{2\rho \omega_{I}}}\left[b_{I}{\bf f}_{I}({\bf r})+b_{I}^{\dag}{\bf f}_{I}^{*}({\bf r})\right],
\label{surface phonon}
\end{equation}
where $b$ and $b^{\dag}$ are the annihilation and creation
operators for the substrate phonons.  The index $I$, like the index $J$ for the
nanoparticle, uniquely specifies a phonon mode for the substrate. ${\bf f}_{I}$ are eigenfunctions
of (\ref{equation of motion}) subject to stress-free boundary conditions at
the $z=0$ plane.

In what follows we will need the spectral density of the isolated substrate which is 
defined as
\begin{equation}
N_{\rm sub}({\bf r},\omega)\equiv-\frac{1}{\pi}{\rm Im}\,D^{zz}_{\rm sub}({\bf r},{\bf r},\omega),
\end{equation}
where $D^{ij}_{\rm sub}({\bf r},{\bf r}',\omega)$ is the Fourier transform of
the retarded phonon Green's function
\begin{equation}
D^{ij}_{\rm sub}({\bf r},{\bf r}',t)\equiv-i\theta(t)\left<\left[u^{i}_{\rm sub}({\bf r},t),u^{j}_{\rm sub}({\bf r}',0)\right]\right>
\label{sub greens function}
\end{equation}
of the substrate.
The spectral density at the free surface of silicon, regarding it as an 
isotropic elastic continuum, was calculated in Appendix B of Ref.~\onlinecite{patton and geller}. There we obtained
\begin{equation}
N_{\rm sub}(\omega)=C_{\rm{Si}}\,\omega,\hspace{.5cm} C_{\rm{Si}}\approx1.4\times 10^{-46}{\rm cm^{2}\,s^2}.
\label{surface spectal density}
\end{equation}

\section{GOLDEN-RULE LIFETIMES}
\label{golden rule lifetimes}
The relaxation rate or inverse lifetime of the perturbed eigenmodes of the 
nanoparticle can be calculated using Fermi's golden rule (setting $\hbar=1$),
\begin{equation}
\tau^{-1}_{J}={2\pi}\sum_{\rm f}|\!\left<{\rm f}|\delta H|{\rm i}\right>\!|^2\delta(\omega_{\rm i}-\omega_{\rm f}),
\end{equation}
where the initial and final states are
\begin{equation}
|{\rm i}\rangle=a^{\dag}_{J}|0\rangle\hspace{.5cm}{\rm and}\hspace{.5cm}|{\rm f}\rangle=b^{\dag}_{I}|0\rangle.
\end{equation}
Using Eqs.~(\ref{perturbation}), (\ref{nano phonon}), and (\ref{surface phonon}) 
leads to\cite{footnote1}
\begin{equation}
\tau^{-1}_{\rm J}=\frac{\pi K^{2}}{2\rho^{2}}\frac{|\Psi^{z}_{J}({\bf r}_0)|^2}{\omega_J}\sum_{I}\frac{|f^{z}_{I}({\bf r}_{0})|^2}{\omega_{J}}\delta(\omega_{J}-\omega_{I}).
\label{lifetime}
\end{equation}
Noting that
\begin{equation}
\sum_{I}|f^{z}_{I}({\bf r}_{0})|^2\delta(\omega_{J}-\omega_{I})=2\rho \omega N_{\rm sub}(\omega),
\end{equation}
we obtain (reinstating factors of $\hbar$) 
\begin{equation}
\tau^{-1}_{J}=\frac{\pi K^2}{\hbar \rho}\frac{N_{\rm sub}(\omega_J)}{\omega_J}|\Psi^{z}_{J}({\bf r}_0)|^2.
\label{golden rule}
\end{equation}
Using $K=1.0\times 10^{4}$  ${\rm erg}\,{\rm cm}^{-2}$ and $\rho=2.3$ ${\rm g}\,{\rm cm}^{-3},$ which are appropriate (see Ref.~\onlinecite{patton and
geller}) for a weak link in Si, 
relaxation rates and Q factors are given in Table \ref{tab:table1} for some
low lying modes. The Q factor is defined here as $\tau$ divided by the period $T$,
\begin{equation}
{\rm Q}\equiv\frac{\tau}{T}=\frac{\hbar \omega_J}{2\pi \gamma},
\end{equation}
where $\gamma\equiv\hbar \tau^{-1}$ is an energy width.

The values of the Q factors we obtain for the low lying modes are incredibly
large, reflecting the fact that the reservoir (substrate) is extremely 
ineffective at absorbing energy at these low frequencies. As we will
discuss below in Section \ref{comparison with experiment}, the lifetimes 
(and Q factors) for the model considered here cannot directly be compared
to the experiment of Ref.~\onlinecite{yang} without accounting for the difference
in sound speeds between a solid Si substrate and a weakly bound nanoparticle
cluster, as well as some other less important modifications.  There we shall
show that the coefficient $C$ in Eq.~(\ref{surface spectal density}) should
be enhanced by about 1000 before making such a comparison, which decreases
the Q factors by this same factor.  However, the Q factors corrected in this
way are still huge, and the good agreement with the observed low-frequency
DOS (see below) suggests that the Q factors of the nanoparticles studied 
experimentally in Ref.~\onlinecite{yang} are also very large.

In Ref.~{\onlinecite{p and g}} we used the golden rule result
(\ref{golden rule}) to estimate the phonon DOS at low energies. This is
achieved by replacing, in accordance with Fermi's golden rule, each
discrete mode in the isolated nanoparticle by a Lorentzian with a width
given by (\ref{golden rule}). (More precisely, this amounts to approximating
the energy-dependent phonon self-energy for each mode
$J$ with its value at $\omega=\omega_{J}$, a procedure often called the
quasiparticle-pole approximation.) However, this procedure is unreliable at low
energies because the actual line-shapes of the broadened modes are 
non-Lorentzian
in the tails.  Nevertheless, we obtained a DOS at 3 ${\rm cm}^{-1}$ that
was only 20 times smaller than that observed.\cite{footnote3}

\begin{table}
\caption{\label{tab:table1}A few representative relaxation rates and Q factors. 
Note that only the $m=0$ spheroidal modes are broadened by the interaction. }
\begin{ruledtabular}
\begin{tabular}{cccc}
$({\rm S},l,m,n)$ & $\omega$\,(rad$\,{\rm s}^{-1}$) & $\tau^{-1}$(${\rm s}^{-1}$) & Q factor\\
\hline
$({\rm S},2,0,1)$ & $3.5\times 10^{12}$ & $2.5\times 10^{-4}$ & $2.2\times 10^{15}$ \\
$({\rm S},1,0,1)$ & $4.9\times 10^{12}$ & $1.0\times 10^{-8}$ & $7.8\times 10^{19}$ \\
$({\rm S},0,0,1)$ & $1.1\times 10^{13}$ & $3.6\times 10^{-4}$ & $4.9\times 10^{15}$ \\
$({\rm S},0,0,2)$ & $1.7\times 10^{13}$ & $1.1\times 10^{-3}$ & $2.5\times 10^{15}$ \\

\end{tabular}
\end{ruledtabular}
\end{table} 
\section{MANY-BODY THEORY OF THE DOS}
\subsection{Local DOS}
\label{local dos}
To leading order in the electron-phonon interaction strength, the electronic 
population relaxation rate due to phonon emission (for example, as measured in 
Ref. \onlinecite{yang}) is given by Fermi's golden rule, which
states the rate (for a deformation potential electron-phonon interaction) is 
proportional to the square of the electron-phonon coupling
strength times
the phonon DOS.  In a translationally invariant system the DOS does not have
any position dependence, but in a nanoparticle one must distinguish between the
``global'' DOS (the DOS relevant for thermodynamics) and the local
eigenfunction-weighted DOS, which is the one that 
determines phonon emission rate.  We will call this position-dependent DOS the
{\it local} DOS, and denote it by $g({\bf r},\omega)$. The precise definition of 
$g({\bf r},\omega)$ will be given below.

From a theoretical point-of-view, the quantity describing the local vibrational
dynamics in the nanoparticle is the (retarded) phonon Green's function
\begin{equation}
D_{\rm R}^{ij}({\bf r},{\bf r}',t)\equiv-i\theta(t)\left<\left[u^{i}({\bf r},t),u^{j}({\bf r}',0)\right]\right>_{H},
\label{nano greens function}
\end{equation}
where
\begin{equation}
\left<\,\cdot\,\right>_{H}\equiv\frac{{\rm Tr}( e^{-\beta H}\,\cdot)}{{\rm Tr}\,e^{-\beta H}},
\end{equation}
and with the Hamiltonian given by
\begin{equation}
H=H_{0}+\delta H.
\end{equation}
Here $H_{0}$ is the Hamiltonian of the isolated nanoparticle and substrate,
\begin{equation}
H_{0}=\sum_{J}\omega_{J}a^{\dag}_{J}a_{J}+\sum_{I}\omega_{I}b^{\dag}_{I}b_{I},
\end{equation}
and, as mentioned in Section \ref{naonoparticle and substrate section}, $\delta H$ is a harmonic spring potential
given in Eq.~(\ref{perturbation}).

In this section the phonon Green's function $D$ always refers to the 
nanoparticle, and the label ``nano'' will be suppressed. The imaginary part 
of the Fourier transform of $D^{ij}_{\rm R}({\bf r},{\bf r},t)$ 
defines the nanoparticle's phonon spectral density
\begin{equation}
N^{ij}({\bf r},\omega)\equiv-\frac{1}{\pi}{\rm Im}\,D^{ij}_{\rm R }({\bf r},{\bf r},\omega).
\label{spectral density}
\end{equation}
For an electron system (or any system of particles), the spectral density 
defined above is precisely the local
DOS. However, because the elasticity equation of motion (\ref{equation of motion}) is second order in time, the spectral density and DOS (both local and 
global)
differ by a factor of $2\rho\omega$. In addition, the vibrational spectral 
density
(\ref{spectral density}) is a tensor, whereas the phonon emission rate probes
some coupling-constant-weighted sum of tensor elements. Because we are 
ascribing the observed reduction in phonon emission (in going from bulk to 
nanoparticle) to a reduction in the local DOS, our results are not sensitive
to the precise way in which a scalar quantity is constructed from the tensor, 
as long
as the same measure is used in both the nanoparticle and bulk.  It will be
most convenient to investigate the trace of the local DOS tensor. Therefore, the 
quantity we calculate in this paper is
\begin{equation}
g({\bf r},\omega)\equiv 2\rho \omega \sum_{i=1}^{3}N^{ii}({\bf r},\omega),
\label{ldos}
\end{equation} 
which we shall refer to as the local DOS. $g({\bf r},\omega)$ characterizes
the number of states per unit energy per unit volume near position ${\bf r}$.
In a bulk material with Debye spectrum, (\ref{ldos}) reduces at low frequency 
to
\begin{equation}
g({\bf r},\omega)=\frac{\omega^2}{2\pi^2}\left(\frac{1}{v_{\rm l}^{3}}+\frac{2}{v_{\rm t}^{3}}\right),
\label{debye dos}
\end{equation}
independent of ${\bf r}$. Eq.~(\ref{debye dos}) is the well-known Debye
formula for the vibrational DOS of a crystal.

The local DOS $g({\bf r},\omega)$ controls the phonon emission rate for an
impurity atom sitting at position ${\bf r}$.  Although the impurity locations 
in a
real nanoparticle are assumed to be random, dopants near the surface are
known to be optically inactive; hence, the experiments (including that of
Ref. \onlinecite{yang}) do not probe the phonon DOS near the nanoparticle
surface.  Therefore, we introduce a particular volume-averaged DOS
\begin{equation}
\bar{g}(b,\omega)\equiv \frac{\int\limits_{r\le b}d^{3}r\,g({\bf r},\omega)}{\frac{4}{3}\pi b^3},
\end{equation}
which characterizes the average $g({\bf r},\omega)$ within a sphere of radius
$b$.
In the limit $b\rightarrow R$, in which case the local DOS is averaged over the full
nanoparticle volume, we obtain the global (or thermodynamic) DOS, which, for
an {\it isolated} nanoparticle, is
\begin{equation}
\bar{g}(R,\omega)=\frac{1}{V}\sum_{J}\delta(\omega-\omega_{J}).
\end{equation}
Physically, we expect $b$ to be somewhere between $R/2$ and $R$.
\subsection{Perturbative Calculation of the Local DOS}
The retarded Green's function (\ref{nano greens function}) for the 
nanoparticle can be obtained by calculating the Euclidean time-ordered 
(or imaginary time) Green's function defined by
\begin{equation}
D^{ij}({\bf r},{\bf r}',\tau)=-\left<Tu^{i}({\bf r},\tau)u^{j}({\bf r}',0)\right>_{H}.
\end{equation}
In the interaction representation,
\begin{equation}
D^{ij}({\bf r},{\bf r}',\tau)=-\frac{\left<Tu^{i}({\bf r},\tau)u^{j}({\bf r}',0)e^{-\int_{0}^{\beta}\delta H(\tau')d\tau'}\right>_{H_0}}{\left<e^{-\int_{0}^{\beta}\delta H(\tau')d\tau'}\right>_{H_0}},
\label{greens function}
\end{equation}
where the expectation values are with respect to $H_{0}$.
By expanding the exponentials to leading order in the perturbation and Fourier 
transforming, (\ref{greens function}) can be written as
\begin{eqnarray}
D^{ij}({\bf r},{\bf r}',\omega)&=&D^{ij}_{0}({\bf r},{\bf r}',\omega)+\sum_{kl}\int D^{ik}_{0}({\bf r},{\bf r}'',\omega)\nonumber\\
&\times&\Pi^{kl}({\bf r}'',{\bf r}''',\omega) D^{lj}_{0}({\bf r}''',{\bf r}',\omega)\,d^{3}r'' d^{3}r''',\nonumber\\
\end{eqnarray}
where 
\begin{equation}
D^{ij}_{0}({\bf r},{\bf r}',\omega)=\sum_{J}\frac{\Psi^{i}_{J}({\bf r})
\Psi^{j*}_{J}({\bf r}')}{2\rho\omega_{J}}\left[\frac{1}{i\omega-\omega_{J}}-\frac{1}{i\omega+\omega_{J}}\right]
\label{unper propagator}
\end{equation}
is the free propagator
and $\Pi^{ij}$ is the leading order self-energy, given at zero temperature by\cite{footnote1}  
\begin{eqnarray}
\Pi^{ij}({\bf r},{\bf r}',\omega)&=&\frac{K^2}{2\rho}\sum_{\rm I}\frac{|f_{\rm I}^{z}({\bf r}_0)|^2}{\omega_{\rm I}}\left[\frac{1}{i\omega-\omega_{\rm I}}-\frac{1}{i\omega+\omega_{\rm I}}\right]\nonumber\\&\times& \delta^{iz}\delta^{jz}\delta({\bf r}-{\bf r}_0)\delta({\bf r}'-{\bf r}_0).
\label{self-energy}
\end{eqnarray}
The ${\bf f}_{\rm I}({\bf r})$ are the substrate eigenfunctions discussed in
Section \ref{naonoparticle and substrate section}, and ${\bf r}_{0}$ is the point at which the nanoparticle is connected to the 
substrate.
Retarded quantities are obtained by analytically continuing $i\omega\to
\omega +i0^{+}$.

To calculate the local DOS (\ref{ldos}) we need to solve the Dyson equation
for the nanoparticle Green's function, written symbolically as
\begin{equation}
D=D_{0}+D_{0}\,\Pi\, D.
\label{dyson}
\end{equation}
A solution to (\ref{dyson}) can be obtained by introducing matrix representations for $D$, $D_{0}$, and $\Pi$, in which case
\begin{equation}
D=(D_{0}^{-1}-\Pi)^{-1}.
\label{symbolic dyson equation}
\end{equation}
The matrix representation we use is defined by
\begin{equation}
O(J,J',\omega)\equiv\sum_{ij}\int_{V} d^{3}r\, d^{3}r' \Psi_{J}^{i*}({\bf r})\Psi_{J'}^{j}({\bf r'})O^{ij}({\bf r},{\bf r}',\omega),
\label{transformation}
\end{equation}
where $O=D$, $D_{0}$, or $\Pi$. In (\ref{transformation}) the integration is
over the volume $V$ of the nanoparticle, and the ${\bf \Psi}_{J}({\bf r})$ are
the nanoparticle eigenfunctions. The inverse transformation is
\begin{equation}
O^{ij}({\bf r},{\bf r}',\omega)=\sum_{JJ'}\Psi_{J}^{i}({\bf r})\,O(J,J',\omega)\,\Psi_{J'}^{j*}({\bf r'}).
\end{equation}
A nanoparticle with a diameter of 10 nm has approximately 8,000 atoms in it, thus,
there are roughly 24,000 acoustic vibrational modes. By knowing the number of 
modes, a Debye energy can be defined: the Debye energy is the energy at which there are 24,000 elasticity-theory modes that lie below in energy. For our 
nanoparticle, the Debye energy is about 240 cm$^{-1}$.  The Debye energy cutoff
truncates the Hilbert space, which leads to finite-size matrices.\cite{footnote4}  This enables 
every mode $J$ of the nanoparticle to be included in the calculation of the 
Green's function (\ref{symbolic dyson equation}).
\section{RESULTS}
\label{results}
In this section we present our results for the phonon DOS in a 10 nm Si nanoparticle, 
obtained by 
solving the Dyson equation (\ref{dyson}) for the phonon Green's function,
as explained above.  As we have discussed, the DOS $g({\bf r},\omega)$, 
defined
in (\ref{ldos}), is a local quantity that varies with position within the
nanoparticle, and, as mentioned in the previous section, the quantity we are 
interested in is $\bar{g}(b,\omega)$, which is $g({\bf r},\omega)$ averaged over
a sphere of radius $b$ centered at the center of the nanoparticle. Because we
have found no  significant dependence of 
$\bar{g}({b},\omega)$ on $b$, for the physically relevant values of $b$ 
($R/2<b<R$), we plot $\bar{g}(R,\omega)$. As stated above in 
Section \ref{local dos},
 $\bar{g}(R,\omega)$ is the global phonon DOS in the nanoparticle.
\begin{figure}
\includegraphics[width=8.5cm]{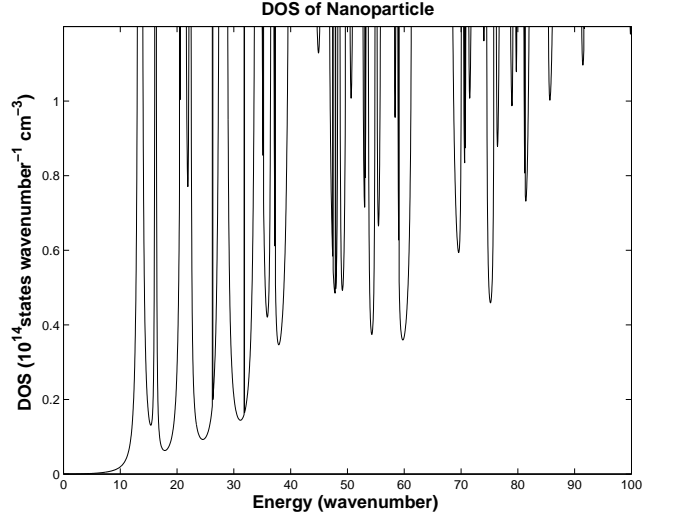}
\caption{\label{highdos} The phonon DOS, given in states per wavenumber per 
cm$^3$, of a 10 nm Si nanoparticle, weakly coupled to a semi-infinite substrate.}
\end{figure}
\begin{figure}
\includegraphics[width=8.5cm]{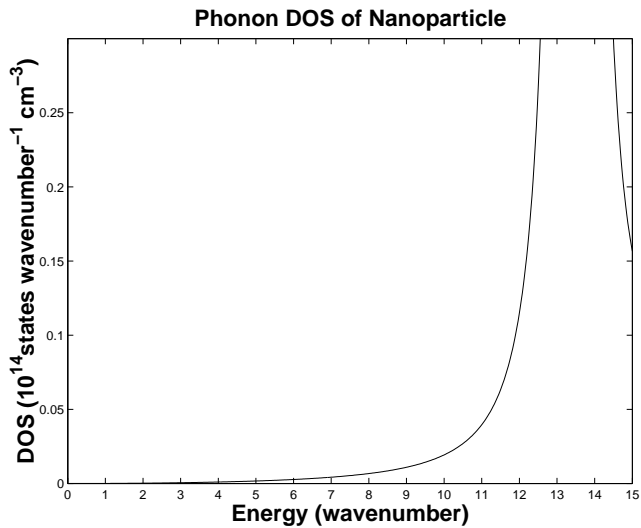}
\caption{\label{lowdos} Vibrational DOS at low energies.}
\end{figure}

\begin{figure}
\includegraphics[width=8.5cm]{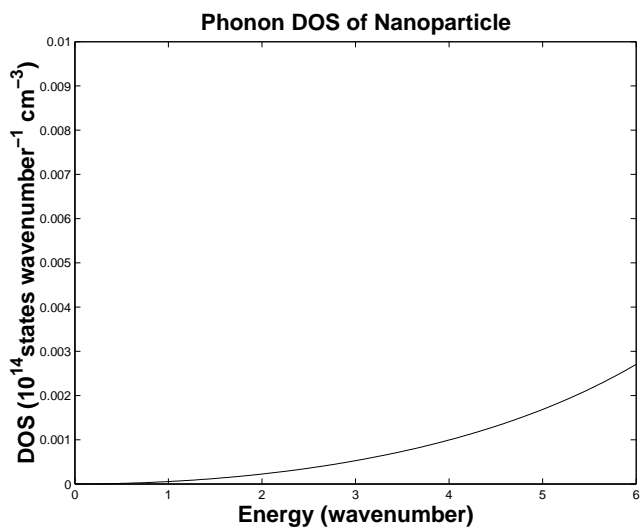}
\caption{\label{lowlowdos} Expanded view of the low-energy DOS. Note that
the DOS vanishes at zero energy, as expected.}
\end{figure}
For simplicity we assume both the nanoparticle and the substrate to 
be made of Si; this allows us to use the surface spectral density (\ref{surface spectal density}) calculated in Appendix B of Ref.~\onlinecite{patton and geller},
where Si is treated as an isotropic elastic continuum with longitudinal and 
transverse sound velocities
\begin{eqnarray}
v_{\rm l}=8.5\times10^5\: {\rm cm}\: {\rm s}^{-1},\nonumber\\
v_{\rm t}=5.9\times10^5\: {\rm cm}\: {\rm s}^{-1},
\end{eqnarray}
and mass density $\rho= 2.3\: {\rm g }\:{\rm cm}^{-3}$. In the final section 
of this paper, where we compare our results to the experiment of Ref.~\onlinecite{yang}, we will introduce an important correction to account for the differences between a solid Si substrate and a nanoparticle cluster.

In Fig.~\ref{highdos} the global DOS $\bar{g}(R,\omega)$ of a 10 nm 
diameter nanoparticle is
given up to 100 cm$^{-1}$.  The modes above 100 cm$^{-1}$ were included in the
calculation, but the long-wavelength approximation of elasticity theory becomes invalid
at high energy. Thus, only the lower part of the spectrum is shown.  Fig.~\ref{lowdos} shows the low-energy phonon DOS up to about 15 cm$^{-1}$.  The large peak on the right side is the well-known Lamb mode. The phonon DOS at 3  
cm$^{-1}$ is approximately $4.9\times 10^{10}$ states per wavenumber per cm$^3$.

\section{COMPARISON WITH EXPERIMENT}
\label{comparison with experiment}
In this section we compare our results with the experiment of Ref.~\onlinecite{yang}, 
where the one-phonon emission rate (and, therefore, the 
phonon DOS at 3 cm$^{-1}$) in a cluster of Y$_{2}$O$_{3}$ nanoparticles was 
observed to be $8.2\times10^{-3}$ times that in bulk 
Y$_{2}$O$_{3}$. In particular, the excited $^{5}{\rm D}_{1}(\rm II)$ state of
Eu$^{3+}$ in the nanoparticles had a phonon-emission lifetime of 27$\mu$s,
compared with a bulk value of 221 ns. In
order to make a comparison of our results to that of the experiment, two
modifications of our calculation have to be performed.

In our model, the cluster of nanoparticles has
been replaced by a solid substrate. However, the spectral density 
(\ref{surface spectal density}) of the substrate, which at long wavelengths, is
determined by the sound speeds and mass density of Si, is much smaller than that 
of
the nanoparticle cluster.  Treating the long-wavelength modes of the cluster
with elasticity theory (or, even simpler, approximating the random cluster by 
an ordered cubic lattice), shows that the spectral density 
(\ref{surface spectal density}) should be replaced by (the subscript ``cl'' 
referring to cluster)
\begin{equation}
N_{\rm cl}(\omega)=C_{\rm cl}\,\omega,
\end{equation}
where\cite{footnote5}
\begin{equation}
C_{\rm cl}=\frac{v_{\rm Si}^{3}\,\rho_{\rm Si}}{v_{\rm cl}^{3}\,\rho_{\rm cl}}C_{\rm Si}.
\label{renormalize}
\end{equation}
Here $v_{\rm cl}$ is a characteristic sound speed in the cluster and 
$\rho_{\rm cl}$ is its mass density.  The $1/v_{\rm cl}^{3}$ dependence in
(\ref{renormalize}) comes from the well-known velocity dependence of the
Debye DOS, and the $1/\rho_{\rm cl}$ factor comes from the definition of
spectral density [see discussion following Eq.~(\ref{spectral density})].
Approximating the cluster by an ordered cubic array with lattice constant
$d$ (the nanoparticle diameter), yields
\begin{equation}
v_{\rm cl}\approx\sqrt{\frac{K}{M}}\,d
\label{reduced velocity}
\end{equation}
and
\begin{equation}
\rho_{\rm cl}\approx\frac{\pi}{6}\rho_{\rm Si},
\label{reduced density}
\end{equation}
where $K$ is the effective spring constant connecting the nanoparticles, given
after Eq.~(\ref{golden rule}), and
\begin{equation} 
M=\frac{4}{3}\pi\rho_{\rm Si}\left(\frac{d}{2}\right)^{3}
\end{equation} 
is the mass of one nanoparticle. Using $d=13$ nm (the mean diameter in Ref.~\onlinecite{yang}), 
we obtain an enhancement factor 
\begin{equation}
\frac{C_{\rm cl}}{C_{\rm Si}}=9.8\times10^2.
\end{equation}
This factor increases the nanoparticle DOS (at frequencies below the Lamb mode) by nearly three orders of magnitude.

There are several other marginally important corrections, most of which will
be ignored, and one that will be included just for completeness: In the
model analyzed above, the nanoparticle was connected to its surroundings by
only a single contact point, whereas a nanoparticle in a cluster most likely has 
more than one connection. As the number of contacts increases, this simply
scales the DOS (away from the peaks) linearly
with the number of contact points. Conservatively, we expect that the 
multiple contact points present in the real system will increase the decay
rate of the nanoparticles' modes, and hence the phonon DOS well below the Lamb
mode, by a factor of two.

The following additional corrections have also been considered and were
found not to be significant (and are not included in our final results):

(i) The actual experiment of Ref.~\onlinecite{yang} was done on an ensemble of
nanoparticles with mean diameter 13 nm and standard deviation of 5 nm. To 
understand the effects of this size distribution, we 
have calculated the DOS at 
3 cm$^{-1}$ averaged over a Gaussian distribution of diameters centered at 10 nm. Even
for very wide distributions (standard deviation up to 8 nm), 
the ensemble
averaged DOS at 3 cm$^{-1}$ is increased by no more than a factor of two.\cite{footnote2} In addition, the  
correction for re-centering the size distribution from 10 nm to the
experimentally observed 13 nm average size, leads to corrections of order unity.

(ii) Our calculations were done for a nanoparticle and substrate made of
Si, while the experiment was done on Y$_{2}$O$_{3}$ nanoparticles, which, of
course, has different mass density and sound velocities.  The differences
in sound speeds and mass only shift around the modes of the nanoparticle 
by a small amount; this has the same effect as a small change in the diameter
of the nanoparticles, which we have found to be negligible.  As for the substrate 
(or more precisely the replacement of
the substrate with a cubic lattice of nanoparticles), the change in mass density
does effect the spectral function (\ref{surface spectal density}) by changing
the velocity (\ref{reduced velocity}) and mass density (\ref{reduced density}), 
but this change is only of order unity.

(iii) The nanoparticles of Ref.~\onlinecite{yang} were immersed in He, either
liquid (for T$<$4.21 K) or gas (T$>$4.21 K). However, the results were found not to 
change through the liquid-gas transition, presumably because of the large 
sound-speed mismatch between superfluid He and Y$_{2}$O$_{3}$.  Therefore, we have
ignored the presence of He in our theory.

(iv) The experiment of Ref.~\onlinecite{yang} was done at temperatures between
1.5 and 10 K (excluding the interval 2.17 to 4.21 K), whereas our calculations assume zero temperature. The effect
of finite temperature is to stimulate phonon emission into the bath (i.e.~substrate or cluster). 
However, this is not important until the Bose distribution function of the bath at the 
Lamb mode frequency (approximately 13 cm$^{-1}$) becomes of order unity, which does not occur 
until the temperature exceeds about 18 K.

Taking into account the first two modifications, and ignoring the rest,
 we obtain a 3 cm$^{-1}$ DOS given by
\begin{eqnarray}
\bar{g}(R,3 {\rm cm}^{-1})&=&4.9\times10^{10}\frac{{\rm states}}{{\rm wavenumber\: cm}^3}
\nonumber\\&\times&\!980\times2\nonumber \\
&=&9.6\times10^{13}\frac{{\rm states}}{{\rm wavenumber\: cm}^3}.
\end{eqnarray}
The DOS of bulk Si is given by
\begin{equation}
\sum_{\lambda}\frac{\epsilon^2}{2\pi^2 \hbar^3 v^3_{\lambda}}=3.9\times10^{15}E^2\frac{{\rm states}}{{\rm wavenumber\: cm}^3},
\end{equation}
where $E$ is the energy in wavenumbers. Thus, the theoretical ratio of the DOS of the 
nanoparticle to that of the bulk material at 3 cm$^{-1}$ is $2.7\times 10^{-3}$. 
As stated above, the experimental ratio of nanoparticle to bulk DOS was found to be 
approximately 
$8.2\times 10^{-3}$. The agreement between our theory and the 
experiment of Ref. \onlinecite{yang} is excellent considering the simplicity and
robustness of our model. We conclude that the low-energy phonon DOS observed
in Ref.~\onlinecite{yang} is consistent with our enviromental broadening mechanism.
\section{CONCLUSIONS}
Motivated by an interesting experiment\cite{yang} measuring the low-energy phonon
DOS in a insulating nanoparticle, we have thoroughly investigated a simplified
model of a single nanoparticle weakly coupled to its environment, a semi-infinite substrate. 
The environmental interactions were found to significantly affect
the DOS at energies below the Lamb mode.

Additionally, we have used the results of our model to predict the effect of
environmental interaction in a cluster of nanoparticles like that studied in
Ref.~\onlinecite{yang}. Although it is necessary to estimate the value of 
several quantities appearing in the model, we believe that we can do this 
accurately enough to obtain a final result that is correct at the order-of-magnitude 
level, with no free parameters.  Because our results for the 3 cm$^{-1}$ DOS is 
only about three times smaller than that observed in Ref.~\onlinecite{yang}, our
broadening mechanism and the resulting phonon spectrum is clearly consistent 
with that experiment.
\section{ACKNOWLEDGMENTS}
This work was supported by the National Science Foundation under CAREER Grant
No.~DMR-0093217, and by the Research Corporation.  Is is a pleasure to thank
Steve Lewis,
Vadim Markel, Richard Meltzer, Mark Stockman, and Ho-Soon Yang for useful
discussions, and Patrick Sprinkle for help with the numerics.  We would especially 
like to thank Bill Dennis for his help with every aspect of this work.


\begin{thebibliography}{99}
\bibitem{footnote1} Throughout this work we use $I$ to label the vibrational modes of the
substrate and $J$ for the nanoparticle.
\bibitem{footnote3} This agreement is fortuitous, however, because we did
not account for the modified sound velocity of the bath.  After this 
modification is made, the golden-rule estimate for the low-frequency DOS
is too large (by a factor of 50 at 3 cm$^{-1}$), as expected.
\bibitem{footnote5} For Si we use the branch-averaged velocity $$v_{\rm Si}=\left(\frac{1}{3}\sum_{\lambda}\frac{1}{v^{3}_{\lambda}}\right)^{-1/3}=6.4\times 10^{5} {\rm cm\,\,s}^{-1}.$$
\bibitem{footnote2} We note, however, that when the Lamb mode frequency becomes resonant with 
the electronic two-level system, the electronic relaxation rate is not simply related to the phonon DOS.
\end{thebibliography}
\end{document}